\documentclass[twocolumn]{svjour3}         
\smartqed  
\usepackage{graphicx}
\journalname{Astrophysics and Space Science}
\begin{document}

\title{Constraint on Cosmological Model with Matter Creation Using
Complementary Astronomical Observations}


\author{Yuan Qiang \and Tong-Jie Zhang \and Ze-Long Yi}


\institute{Yuan Qiang \at Department of Astronomy, Beijing Normal University,
Beijing 100875, P.R.China \\
Key Laboratory of Particle Astrophysics, Institute of High Energy Physics,
Chinese Academy of Sciences, P.O.Box 918-3,Beijing 100049, P.R.China\\
     \email{yuanq@mail.ihep.ac.cn}            
           \and
        Tong-Jie Zhang \at Department of Astronomy, Beijing Normal University,
Beijing 100875, P.R.China \\
     \email{tjzhang@bnu.edu.cn}
           \and
        Ze-Long Yi \at Department of Astronomy, Beijing Normal University,
Beijing 100875, P.R.China\\
}

\date{Received: date / Accepted: date}

\maketitle

\begin{abstract}
The universe with adiabatic matter creation is considered. It is
thought that the negative pressure caused by matter creation can
play the role of a dark energy component, and drive the
accelerating expansion of the universe. Using the Type Ia
supernovae (SNe Ia) data, the observational Hubble parameter data, the
Cosmic Microwave Background (CMB) data and the Baryonic Acoustic
Oscillation (BAO) data, we make constraints on the cosmological
parameters, assuming a spatially flat universe. Our results show
that the model with matter creation is consistent with the SNe Ia
data, while the joint constraints of all these
observational data disfavor this model. If the cosmological constant
is taken into account, a traditional model without matter creation
is favored by the joint observations.
\end{abstract}

\section{Introduction}
\label{intro}
A great encouraging development in modern cosmology is the
discovery of the accelerating expansion of the universe through
observations of distant Type Ia
Supernovae \cite{Riess98,Perlmutter99}. The
anisotropy of the cosmic microwave background (CMB) results from
balloon and ground
experiments \cite{Miller99,Bernardis00,Hanany00,Halverson02,Mason03,Benoit03}
and recent WMAP \cite{Spergel03} observation confirmed the
result from SNe Ia and favored a spatially flat universe. It is
well known that all known types of matter with positive pressure
generate attractive forces and decelerate the expansion of the
universe. The discovery from SNe Ia and CMB indicate the existence
of a new component with negative pressure, which is now generally
called dark energy.

The simplest one of dark energy is a cosmological
constant \cite{Weinberg89,Carroll92,Ostriker95}.
An explanation of the cosmological constant
is the vacuum energy, however, it is 120 orders of magnitude
smaller than the naive expectation from quantum field theory.
Bothering physicists much, other types of dark energy are
proposed, such as
quintessence \cite{Ratra88,Coble97,Caldwell98},
which is described in terms of a cosmic
scalar field $\phi$; or other modified cosmological models are
discussed, such as the Cardassian expansion model which
investigates the acceleration of the universe by a modification to
the Friedmann equations
\cite{Freese02,Zhu02}, the brane world
model which explain the acceleration through the fact that general
relativity is formulated in 5 dimensions instead of the usual
4 \cite{Randall99,Deffayet02,Avelino02} and so on.

All of these dark energy cosmological models are based on the Big
Bang cosmology. A model with adiabatic matter creation was
proposed firstly in order to interpret the cosmological entropy
and solve the big-bang singularity problem \cite{Prigogine89}.
The basic idea is to modify the usual energy conservation law in
open system in the framework of cosmology, which adds a balance
equation for the number density of the created particles to the
dynamic equations of the universe. Nevertheless, after the
discovery of the accelerating expansion of the universe, this
model was reconsidered to explain it and got some unexpected
results. The matter creation pressure $p_c$, which is negative
as pointed out several decades ago by Zel'dovich \cite{Zeldovich70},
might play the role of a dark energy component and lead to the
accelerating expansion of the universe. Lima \& Alcaniz
tested the model without cosmological constant through the
lookback time-redshift relation, luminosity distance-redshift
relation, angular size-redshift relation and the galaxy number
counts-redshift relation \cite{Lima99,Alcaniz99}. It was shown that this
model was consistent with the observational accelerating expansion of the
universe, and could also alleviate the conflict between the age of
the universe and the age of the oldest globular clusters.
Zimdahl et al. employed the SNe Ia data to test the matter creation
scenario and also got the result of accelerating expansion \cite{Zimdahl01}.
Freaza et al., however, based on the observational SNe Ia data and the
simulated Supernova Acceleration Probe (SNAP) data, showed the
matter creation mechanism was not favored to explain the cosmic
acceleration \cite{Freaza02}.

In this paper, we use recently 186 SNe Ia
sample \cite{Riess04}, combined with the observational
$H(z)$ data from the differential age measurements of galaxies
\cite{Simon05}, the CMB and BAO data \cite{Wang06,Eisenstein05},
to test the cosmological model with matter
creation and make constraints of the parameters. As a
comparison, the model with both matter creation and cosmological
constant is also examined. This paper is organized as follows: we
present the basic cosmological equations of the universe with
adiabatic matter creation in Sec.~\ref{sec:1}. In Sec.~\ref{sec:2}
we give a brief introduction to the observational data, and
the results and discussion are given in Sec.~\ref{sec:3}.

\section{The Cosmological Basic Equations with Adiabatic Matter
Creation}
\label{sec:1}
The Robertson-Walker(RW) metric describing the
space-time of the universe is
\begin{equation}
{\rm d}s^2=-{\rm d}t^2+a^2(t)[\frac{{\rm d}r^2}{1-kr^2}+r^2{\rm
d}\theta^2+r^2\sin^2\theta{\rm d}\phi^2], \label{RW-metric}
\end{equation}
where $r,\theta,$ and $\phi$ are dimensionless comoving
coordinates, $k=0,\pm1$ represent the curvature of the spatial
section and $a(t)$ is the scale factor. Using the Einstein field
equation, we can acquire the equations to describe the dynamic
behavior of the universe, namely the Friedmann equations
\begin{eqnarray}
\frac{\dot{a}^2}{a^2}&=&\frac{8\pi
G}{3}\rho+\frac{\Lambda}{3}-\frac{k}{a^2},
\label{Friedmann1}\\
\frac{\ddot{a}}{a}&=&-\frac{4\pi
G}{3}[\rho+3(p+p_c)]+\frac{\Lambda}{3}, \label{Friedmann2}
\end{eqnarray}
where $\rho=\rho_M+\rho_R$ is the energy density(matter and
radiation), $p$ and $\Lambda$ are the thermal pressure and
cosmological constant respectively. $p_c$ is the matter creation
pressure that takes the following form \cite{Calvao92}
\begin{equation}
p_c=-\frac{\rho+p}{3nH}\psi, \label{pc}
\end{equation}
where $H\equiv \dot{a}/a$ is the Hubble parameter and $\psi$ is
the matter creation rate. In models with adiabatic creation, the
balance equation for the particle number density
$n$ is \cite{Calvao92,Lima92}
\begin{equation}
\frac{\dot{n}}{n}+3\frac{\dot{a}}{a}=\frac{\psi}{n}.
\label{n_balance}
\end{equation}
We take the form of matter creation rate
as \cite{Lima99}
\begin{equation}
\psi=3\beta nH, \label{create_rate}
\end{equation}
where the parameter $\beta$ is defined on the interval $[0,1]$ and
assumed to be constant. Matter and radiation correspond to
$\beta_M$ and $\beta_R$ respectively. Together with the equation
of state (EOS)
\begin{equation}
p=w\rho, \label{EOS}
\end{equation}
the equation system becomes complete.

From Eqs.(\ref{Friedmann1})and(\ref{Friedmann2}), we can get
\begin{equation}
p+p_c=-\frac{{\rm d}(\rho a^3)}{{\rm d}(a^3)}. \label{conserv}
\end{equation}
Combining Eqs.(\ref{pc}),(\ref{EOS})and(\ref{conserv}), it is easy
to find
\begin{equation}
\left\{
\begin{array}{lll}
w=0, & \rho_M=\rho_{M0}(\frac{a_0}{a})^{3(1-\beta_M)} & {\rm for
\ matter}\\
w=1/3, & \rho_R=\rho_{R0}(\frac{a_0}{a})^{4(1-\beta_R)} &
{\rm for\ radiation}\\
w=-1, & \rho_{\Lambda}=\rho_{\Lambda0} & {\rm for\ cosmological}\\
      & & {\rm constant}
\end{array}\right..
\label{density}
\end{equation}
Using Eq.(\ref{density}) and noting that $1+z=a_0/a$, we can
rewrite Eq.(\ref{Friedmann1}) in the form of the Hubble parameter
\begin{equation}
H^2=H_0^2E^2(z), \label{hubble}
\end{equation}
where $H_0$ is the Hubble constant, and
\begin{eqnarray}
E^2(z)&=&\Omega_M(1+z)^{3(1-\beta_M)}+\Omega_R(1+z)^{4(1-\beta_R)}
\nonumber \\
      &+&\Omega_{\Lambda}+\Omega_k(1+z)^2 \label{ez}
\end{eqnarray}
represents the expansion rate, in which
$\Omega_M,\Omega_R,\Omega_{\Lambda}$ and $\Omega_k$ are the matter
density, radiation density, cosmological constant and spatial
curvature parameters at present.

The deceleration parameter $q$ is defined as
\begin{equation}
q=-\frac{a\ddot{a}}{\dot{a}^2}.
\end{equation}
Using Eqs.(\ref{Friedmann1}),(\ref{Friedmann2}) and
(\ref{density}), the deceleration parameter $q$ reduces to
\begin{eqnarray}
q(z)&=&\frac{1}{E^2(z)}[\frac{1-3\beta_M}{2}\Omega_M(1+z)^{3(1-\beta_M)}
\nonumber \\
    &+&(1-2\beta_R)\Omega_R(1+z)^{4(1-\beta_R)}-\Omega_{\Lambda}].
\label{qz}
\end{eqnarray}
In the following discussion, we consider a spatially flat
universe\cite{Spergel03}, and neglect the radiation
term for its extremely small value today, that is,
$\Omega_k=\Omega_R=0$, so $\Omega_M+\Omega_{\Lambda}=1$.

If there is no cosmological constant, that is,
$\Omega_{\Lambda}=0,\Omega_M=1$, Eq.(\ref{ez}) can be rewritten as
\begin{equation}
E^2(z)=(1+z)^{3(1-\beta)}, \label{ez1}
\end{equation}
where $\beta$ is $\beta_m$, and $q(z)$ can be simplified as
\begin{equation}
q(z)=\frac{1-3\beta}{2},\label{qz1}
\end{equation}
which shows that $q$ is independent of redshift. From
Eq.(\ref{qz1}) we know that, if the universe is accelerating
expanding, that is, $q_0<0$, $\beta$ needs to be greater than
$1/3$.

\section{The Observational Data}
\label{sec:2}
SNe Ia are thought to be standard candles and
can be used as distance probes. The theoretical prediction for
luminosity distance of an astronomical object in a spatially flat
universe is
\begin{equation}
d_L=\frac{c(1+z)}{H_0}\int_0^z\frac{{\rm d}z}{E(z)}, \label{dL}
\end{equation}
where $E(z)$ is defined in Eq.(\ref{hubble}). The distance modulus
is
\begin{eqnarray}
\mu_{th}=5\lg\frac{d_L}{Mpc}+25=42.38+5\lg[\frac{1+z}{h}
\int_0^z\frac{{\rm d}z}{E(z)}], \label{mu}
\end{eqnarray}
where $h=H_0/(100\,km/s/Mpc)$. The $\chi^2$ parameter for SNe Ia
is
\begin{equation}
\chi^2_{SN}=\sum_{i}\frac{[\mu_{th}(z_i;h,\Omega_M,\beta)-
\mu_{SN}(z_i)]^2}{\sigma_{SN}(z_i)}, \label{chi2SN}
\end{equation}
where $\mu_{SN}(z_i)$ is the observed distance modulus of the SN with
redshift $z_i$ and $\sigma_{SN}(z_i)$ is the observational error.

We also include the observational Hubble parameter $H(z)$ in this
work. The Hubble parameter depends on the differential age of the
universe in this form
\begin{equation}
H(z)=-\frac{1}{1+z}\frac{{\rm d}z}{{\rm d}t},\label{hz}
\end{equation}
which provides a direct measurement for $H(z)$ through a
determination of ${{\rm d}z}/{{\rm d}t}$. Using the differential
ages of passively evolving galaxies determined from the Gemini Deep
Deep Survey (GDDS), Simon et al. determined a set of observational
$H(z)$ data in the redshift range $0\sim1.8$ 
\cite{Simon05}. These observational
$H(z)$ data have been used to constrain the dark energy potential
and its redshift dependence by Simon et al. \cite{Simon05}. Various
cosmological models were tested using this data set in recent several
years. Yi \& Zhang used them to analyze the holographic dark energy models 
and showed that the fitting results did not conflict with some other
independent cosmological tests \cite{Yi07}. Samushia \& Ratra used
the data set to constrain the parameters of $\Lambda$CDM, XCDM and
$\phi$CDM models. The constraints are consistent with those derived 
from SNe Ia \cite{Samushia06}. Wei \& Zhang tested a series of other 
cosmological models with interaction between dark matter and dark energy
using this data set. \cite{Wei07}. In one word, the
observational $H(z)$ data are demonstrated to be an effective 
complementarity to other cosmological probes. 

For the Hubble parameter $H(z)$ data, we have
\begin{equation}
\chi^2_H=\sum_k\frac{[H_{th}(z_i;h,\Omega_M,\beta)-H_{ob}(z_i)]^2}
{\sigma_{H}(z_i)}.\label{chi2H}
\end{equation}

The model-independent shift parameter $R$, which can be derived
from CMB data, is defined as
\begin{equation}
R=\sqrt{\Omega_M}\int_0^{z_r}\frac{{\rm d}z}{E(z)},\label{R}
\end{equation}
with $z_r=1089$ the redshift of recombination. From the three-year
result of WMAP, Wang \& Mukherjee estimated $R=1.70\pm0.03$ \cite{Wang06}.

Using a large spectroscopic sample of luminous red galaxies from
Sloan Digital Sky Survey (SDSS), Eisenstein et al.
successfully found the acoustic peaks in the bayonic matter
anisotropy power spectrum, described by the model-independent $A$
parameter
\begin{equation}
A=\sqrt{\Omega_M}\left
[\frac{1}{z_1E^{1/2}(z_1)}\int_0^{z_1}\frac{{\rm d}z}{E(z)}\right
]^{2/3},\label{A}
\end{equation}
with $z_1=0.35$ the redshift at which the acoustic scale has been
measured \cite{Eisenstein05}. Eisenstein et al. suggested the value
of $A$ parameter as $A=0.469\pm0.017$.

The model parameters $h,\Omega_M$ and $\beta$ can be determined
through the $\chi^2$ minimization method. The combined $\chi^2$
can be written as
\begin{equation}
\chi^2=\chi^2_{SN}+\chi^2_H+\frac{(R-1.70)^2}{0.03^2}+
\frac{(A-0.469)^2}{0.017^2}. \label{chi2}
\end{equation}
By minimizing $\chi^2$ we can get the best fitting values of the
parameters.

\section{Results and Discussion}
\label{sec:3}
\begin{figure}
\centering
\includegraphics[width=0.45\textwidth]{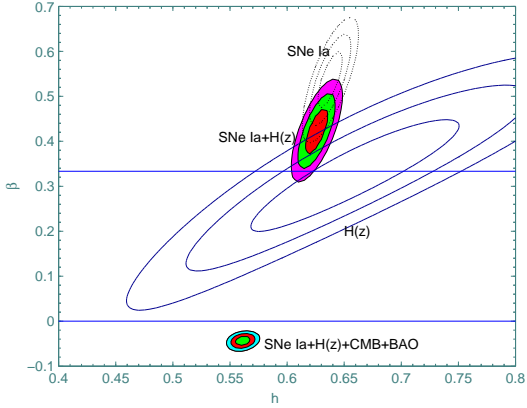}
\caption{Confidence regions of 1, 2 and 3
$\sigma$ (from inside to outside) for the two parameters $h$ and
$\beta$, for different observational data set as labelled in the
figure. The horizon lines are of $\beta=1/3$ and $0$. \label{fig1eps}}
\end{figure}

Using the observations of SNe Ia, Hubble parameter $H(z)$,
CMB and BAO, we test the cosmological model with adiabatic matter
creation, assuming a spatially flat universe.
We show in Fig. \ref{fig1eps} the confidence regions of parameters
$h$ and $\beta$ of this model, for different sets of observational
data. The best fitting results of parameters are listed in Table
\ref{table1}. We can see from this figure that, for the SNe Ia data,
$\beta$ is greater that $1/3$ at $3\sigma$ level, which means that the
universe is accelerating expanding. The results of $H(z)$ data are
a little different from which of SNe Ia. The best fitting value
of $\beta$ is $0.33$ for $H(z)$ data. Because of the smallness of
statistical sample of $H(z)$ data, the error bars of parameters
are relatively large and the $H(z)$ data fail to constrain the
parameters strictly enough. $H(z)$ data cannot constrain the model
parameters very well. The joint constraints of all these data (SNe
Ia + H(z) + CMB + BAO) are plotted in the same figure. We find
$\beta<0$ at $3\sigma$ level, with the best fitting value $\beta=
-0.04$. A negative value of $\beta$ means matter vanishing instead
of matter creation, which is forbidden thermodynamically
\cite{Calvao92}. We draw the conclusion that this model
is consistent with SNe Ia observations, but is disfavored by
the joint analysis from different observations.

\begin{table}
\caption{Best fitting values of
$h$ and $\beta$ of the matter creation model without the cosmological
constant.}
\centering
\begin{tabular}{cccc}
\hline\noalign{\smallskip}
Test & $h$ & $\beta$ & $\chi^2/d.o.f$  \\
\noalign{\smallskip}\hline\noalign{\smallskip}
  SNe Ia                & $0.64$ & $0.54$ & $239.97/184$  \\
  H(z)                  & $0.66$ & $0.33$ & $8.50/7$  \\
  SNe Ia+H(z)           & $0.62$ & $0.42$ & $269.40/193$  \\
  SNe Ia+H(z)+CMB+BAO   & $0.56$ & $-0.04$ & $753.71/195$  \\
\noalign{\smallskip}\hline
\end{tabular}
\label{table1}
\end{table}

\begin{figure}
\centering
\includegraphics[width=0.45\textwidth]{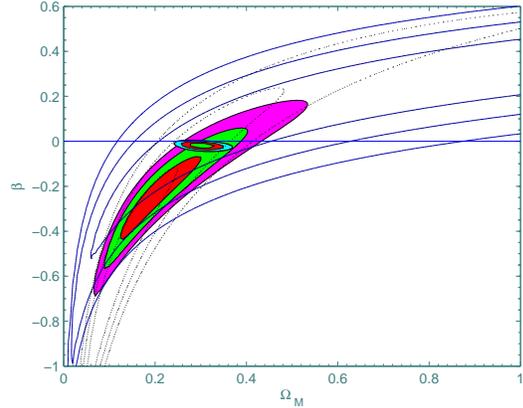}
\caption{Confidence regions for the two parameters $\Omega_M$ and
$\beta$,for the cosmological model with both the matter creation
and the cosmological constant. Parameter $h$ has been marginalized.
The lines are coded as Fig. \ref{fig1eps}.}\label{fig2eps}
\end{figure}

\begin{table}
\caption{Best fitting values of
$\Omega_M$ and $\beta$ of the matter creation model with the cosmological
constant ($h$ marginalized).}
\centering
\begin{tabular}{cccc}
\hline\noalign{\smallskip}
Test & $\Omega_M$ & $\beta$ & $\chi^2/d.o.f$  \\
\noalign{\smallskip}\hline\noalign{\smallskip}
  SNe Ia                & $0.10$ & $-0.66$ & $237.64/184$  \\
  H(z)                  & $0.91$ & $0.31$ & $13.94/7$  \\
  SNe Ia+H(z)           & $0.19$ & $-0.26$ & $250.87/193$  \\
  SNe Ia+H(z)+CMB+BAO   & $0.30$ & $-0.02$ & $256.42/195$  \\
\noalign{\smallskip}\hline
\end{tabular}
\label{table2}
\end{table}

One can expect that the negative pressure of both the cosmological
constant and the matter creation may jointly generate the
acceleration of the universe. We also test the model with a
cosmological constant in order to
give a comparison. After marginalizing parameter $h$, we show the
confidence regions of parameters $\Omega_{M}$ and $\beta$ in Fig.
\ref{fig2eps} and the best fitting parameters in Table \ref{table2}
respectively. The cosmological constant $\Omega_{\Lambda}=1-\Omega_{M}$.
For the SNe Ia and $H(z)$ data, the constraints of model parameters
are not very strong. The statistical uncertainties of model parameters
are so large that we cannot give any convincible conclusions. While
the joint constraints of these 4 kinds of observations show that
$\beta$ tends to be $zero$, which reduces to the familar $\Lambda$CDM
cosmology. The best fitting result $\Omega_M=0.30\pm 0.02$ is also
consistent with other studies of the traditional $\Lambda$CDM model
\cite{Riess04,Spergel03}. We also notice that our result is very
similar with \cite{Freaza02}. However, in their treatment, a small
SNe Ia sample (16 low-redshift and 38 high-redshift supernovae from
Perlmutter et al. \cite{Perlmutter99}) could not give strong constrains
on the model parameters. They drew the conclusion by adopting a
Gaussian prior of $\Omega_M=0.27\pm0.06$. In our work, we give the
constraints from joint astronomical observations directly.

In summary, the idea that a negative
matter creation pressure may play the role of dark energy and
drive the accelerating expansion of universe, seems not favored
by the combined observations. While for the model with both the
adiabatic matter creation and cosmological constant, the joint
constraints tend to give $\beta\simeq 0$, which means
no matter creation, and the model becomes the traditional $\Lambda$CDM
one.

\begin{acknowledgements}
We would like to thank Yue Bin, Dr.Peng-Jie
Zhang, Li Chen and Zong-Hong Zhu, for their friendly
discussion. This work was supported by the National Science
Foundation of China (Grants No.10473002 and 10273003), the
Scientific Research Foundation for the Returned Overseas Chinese
Scholars, State Education Ministry and the Research Foundation for
Undergraduate of Beijing Normal University.
\end{acknowledgements}

\bibliographystyle{natbib}


\end{document}